# Absence of polar order in LuFe$_2$O$_4$

A. Ruff[1], S. Krohns[1], F. Schrettle[1], V. Tsurkan[1,2], P. Lunkenheimer[1,a], and A. Loidl[1]

[1]Experimental Physics V, Center for Electronic Correlations and Magnetism, University of Augsburg, 86135 Augsburg, Germany

[2]Institute of Applied Physics, Academy of Sciences of Moldova, MD 2028, Chisinau, R. Moldova

**Abstract.** LuFe$_2$O$_4$ often is considered as a prototypical multiferroic with polar order arising from the electronic degrees of freedom only ("electronic ferroelectricity"). In the present work, we check the intrinsic nature of the dielectric response of this material by performing dielectric measurements of polycrystalline samples with different types of contact materials and with different grain sizes. In addition, frequency-dependent measurements of the electric-field dependent polarization are provided. The obtained results unequivocally prove that the reported colossal dielectric constants in LuFe$_2$O$_4$, which were interpreted in terms of electronic ferroelectricity, are of non-intrinsic surface-related origin. The intrinsic dielectric properties of this material show no indications of any ferroelectric order and, thus, LuFe$_2$O$_4$ is not multiferroic. Its intrinsic dielectric constant is close to 20 and its dielectric loss is dominated by charge transport via variable range hopping.

PACS numbers: 77.22.Ch, 75.85.+t, 77.84.Bw, 72.20.Ee

## 1 Introduction

The interest in multiferroics, specifically in systems combining magnetism and ferroelectricity, gained enormous impetus after the discovery of spin-driven ferroelectrics by Kimura *et al*. [1]. This new class of magnetic transition-metal oxides naturally opens the possibility to design materials with strong magneto-electric coupling, i.e. which allow to induce magnetic response by electric fields and vice versa. Soon after this breakthrough it also became clear that in systems with non-integer valence of the transition-metal ions, like doped manganites or magnetite, ferroelectricity may occur via complex charge ordering phenomena [2,3]. While the theoretical concepts are elegant and rather straightforward, from an experimental point of view the situation is much less settled and more complicated. All these systems, which we further on will call electronic ferroelectrics, are usually characterized by rather high conductivity, sometimes even in the charge ordered state. Thus, the detection of their intrinsic dielectric response can be expected to be hampered by extrinsic, so-called Maxwell-Wagner (MW) effects, which stem from charge accumulation and/or depletion layers at the sample surface or at internal barriers like grain boundaries. MW effects can formally be described by simple electronic equivalent circuits, leading to a dielectric response identical to Debye relaxation phenomena, and in many cases yield colossal values of the dielectric constants which by no means point towards polar order [4,5,6]. Internal or external barrier layers can also be responsible for the observation of pyrocurrents and "banana-like" ferroelectric hysteresis curves [7,8].

Focusing on the existing experimental situation, three magnetic materials with electronic ferroelectricity have been reported so far: LuFe$_2$O$_4$ (LFO) [9], Fe$_3$O$_4$ [10,11,12] and (PrCa)MnO$_3$ [13,14]. In the latter compound, even at low temperatures the conductivity is high (of order 10 - 100 $\Omega$cm) and the dielectric results cannot be easily interpreted assuming a polar phase transition. Thus, there is only indirect experimental evidence of a polar state in this material. For LFO, the documented dielectric response, as well as the pyrocurrent results [9] are no proof of a polar state and could as well stem from extrinsic MW effects. A critical review of the existing literature shows that in these systems, ferroelectric hysteresis curves so far have only been observed in magnetite at low temperatures [10,11,12] and from detailed broadband dielectric spectra, it has been concluded that the low temperature polar state of magnetite has to be characterized as relaxor ferroelectric [12].

But also from a structural point of view the charge order scenarios recently have been revisited, allowing new insights into the charge ordered low-temperature states: With highly refined structural techniques, the charge order in magnetite has been identified as localized electron distribution over three iron-site units (trimerons), supporting a low-temperature polar state [15]. On the other hand, single-crystal x-ray diffraction data on LFO imply non-polar bilayers, inconsistent with charge-order induced ferroelectricity [16]. Indeed, recent dielectric work on poly [17] and single crystals [18] do indicate that the apparent polar state reported earlier corresponds solely to extrinsic MW effects. In view of a variety of supporting models for polar order in LFO, it seems vital to arrive at a final conclusion. Analyses of complex impedance plots [17,19] and fits with equivalent circuits [4,5,6,17,18] are very useful to help revealing a non-intrinsic origin of dielectric results and this indeed was recently carried out in LFO [17,18]. However, for a final proof and to definitely clarify the origin of the observed colossal values of the dielectric constant in LFO (contacts or grain boundaries), dielectric measurements after a variation of contact types and grain sizes are essential [4]. Hence, we have investigated the dielectric properties of polycrystalline LFO as function of contact material and tempering time to ultimately determine the nature of the dielectric low-temperature state in LFO.

---
[a] e-mail: peter.lunkenheimer@physik.uni-augsburg.de

## 2 Experimental details and sample characterization

Polycrystalline samples of $LuFe_2O_4$ were prepared by a two-step synthesis via solid-state reactions. Binary oxides $Lu_2O_3$, $Fe_2O_3$, and FeO of 99.99 % purity were used as starting materials for the synthesis. At the first step, the ternary $LuFeO_3$ phase was synthesized at ambient conditions at 1000°C. At the second step, reground $LuFeO_3$ mixed with FeO was pressed into pellets and sintered three times for one week in evacuated quartz ampoules with intermediate regrinding after each sintering procedure. The vacuum condition is necessary to prevent the oxidation of iron with the final product of $LuFeO_3$. At the second step, the sintering temperature was 1000°C, and after the second and third regrinding, the sintering temperature was increased to 1100°C. The second $LuFe_2O_4$ sample was additionally sintered for two weeks at 1100°C to increase the grain size. X-ray diffraction with subsequent Rietveld analysis documented a $LuFe_2O_4$ phase of $R\bar{3}m$ symmetry with lattice constants $a$ = 0.3433(1) nm and $c$ = 2.5227(4) nm without impurity phases within experimental uncertainty.

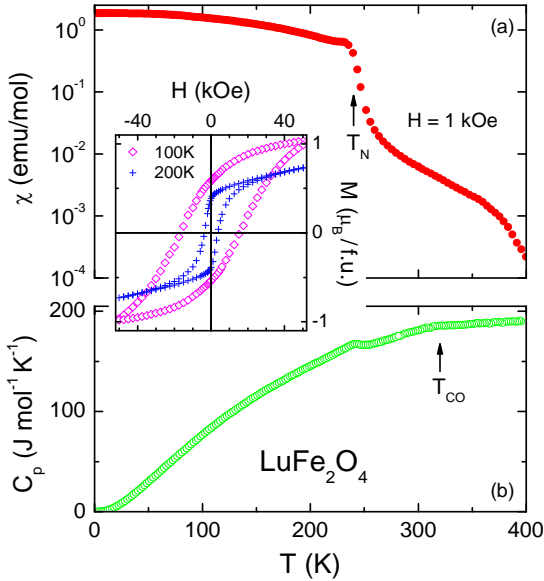

**FIG. 1.** Temperature dependence of the magnetic susceptibility (measured at 1 kOe) (a) and of the specific heat (b) of LFO. The inset shows the magnetization vs. external field taken at two temperatures below $T_N$.

The samples have been characterized by magnetic susceptibility using a SQUID magnetometer (Quantum Design MPMS-5) and by heat capacity experiments (Quantum Design PPMS). The results covering temperatures from 2 K up to 400 K are shown in Fig. 1. The magnetic susceptibility in the upper frame as measured at 1 kOe shows a step-like increase close to 240 K, indicative for the onset of ferrimagnetic order due to a specific arrangement of charge ordered $Fe^{2+}$ and $Fe^{3+}$. The inset in Fig. 1 shows magnetic hysteresis curves at 100 and 200 K and provides clear experimental evidence for a ferromagnetic moment of the order of 1 $\mu_B$ per formula unit. The origin of the significant temperature variation of the shape of the hysteresis is unclear at present. It may result from a strong magnetic anisotropy and qualitatively similar behavior was also found for single crystals [20,21]. Compared to single crystals, the coercitivity of the polycrystalline samples is strongly enhanced due to grain boundary and surface related effects.

The magnetic transition is accompanied by a clear anomaly close to 240 K in the temperature-dependent heat capacity $C(T)$, documented in Fig. 1(b). The transition temperature again is close to 240 K. The charge order transition at $T_{co}$ is indicated as minor bump in $C(T)$ and as a point of inflection in the magnetic susceptibility. The charge-order transition occurs close to 320 K, in accordance with reports in literature [16,18,22]. Neither in heat capacity nor in magnetic susceptibility we find evidence for a further low-temperature transition close to 170 K where significant broadening of the magnetic Bragg reflections was reported [23].

For the dielectric measurements, silver paint or sputtered silver contacts were applied at opposite sides of the pellets. The dielectric constant and conductivity were determined in a broad frequency range between 0.1 Hz and 2.3 GHz using a frequency-response analyzer (Novocontrol α-analyzer, 0.1 Hz - 1 MHz), autobalance bridges (Hewlett-Packard 4284A and Agilent E4980A, 20 Hz - 1 MHz), and an impedance analyzer (Agilent E4991A, 1 MHz - 2.3 GHz) applying voltages of 0.5 - 1 V [24,25]. Additional non-linear polarization measurements were performed with a ferroelectric analyzer (aixACCT TF2000). For sample cooling and heating, a closed-cycle refrigerator system and a Nitrogen-gas cryostat were used.

## 3 Results and discussion

Figure 2 shows the temperature dependence of the dielectric constant and conductivity of LFO for various frequencies of the exciting ac field. The figure includes the combined results from several measurement runs, covering a temperature range from 20 - 500 K and frequencies of $10^{-1}$ - $2.3 \times 10^9$ Hz. The same sample with sputtered silver contacts was used in all runs. In agreement with earlier work [9,17,18,26], at high temperatures and low frequencies, dielectric constants of colossal magnitude are observed [Fig. 2(a)]. With decreasing temperature, $\varepsilon'(T)$ shows two consecutive steplike drops and finally assumes values of about 20. These drops of $\varepsilon'$ strongly shift to lower temperatures with decreasing frequency, which is the typical signature of relaxational processes. Remarkably, at its upper plateau $\varepsilon'(T)$ reaches levels larger than $10^5$, exceeding all previously reported values of $\varepsilon'$ in this material [9,17,18,26].

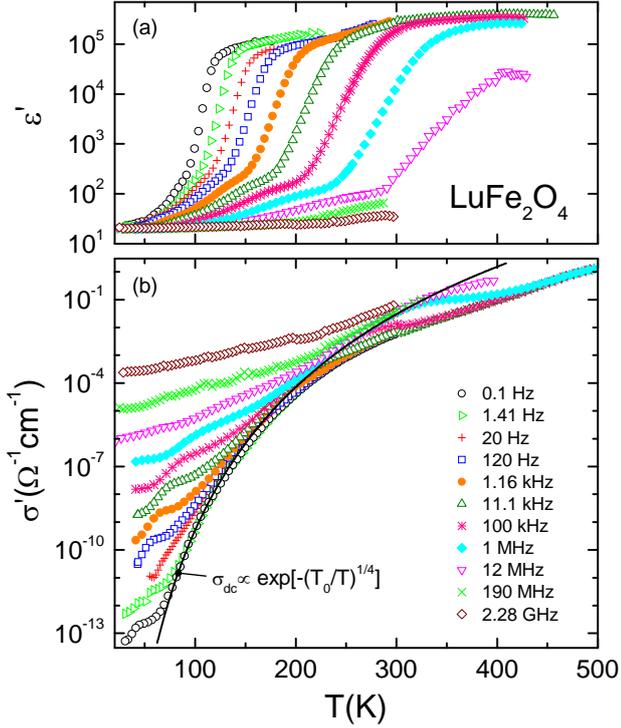

**FIG. 2.** Temperature dependence of the dielectric constant (a) and conductivity (b) of LFO measured at various frequencies. The contacts were prepared by sputtered silver. The line in (b) provides an estimate of the intrinsic dc-conductivity. It was calculated assuming VRH behavior.

For relaxation processes, peaks should show up in the temperature dependence of the dielectric loss $\varepsilon''(T)$ or of the conductivity $\sigma'(T) \propto \varepsilon''(T)\nu$. They should be located at frequencies comparable to those of the steps in $\varepsilon'(T)$. As revealed in Fig. 2(b), in the present case those peaks are superimposed on the strongly temperature dependent charge-transport contribution of the semiconducting material. This superposition gives rise to shoulders, which shift to higher temperatures with increasing frequency. These shoulders are especially well pronounced for the low-temperature relaxation (e.g., at about 80 K for the 100 kHz curve). For the high-temperature relaxation (e.g., at about 275 K for 100 kHz) they are rather smeared out due to the stronger dc-conductivity contribution at high temperatures.

To reveal the origin of the observed relaxational processes and of the high magnitude of the dielectric constant of LFO, in Fig. 3 the temperature dependence of the dielectric constant and conductivity, measured with two different contact types are presented: While the lines were obtained with silver-paint contacts, the symbols represent the results for sputtered-silver contacts, applied to the same sample after removal of the silver paint. Remarkably, for silver-paint contacts the absolute values of $\varepsilon'(T)$ at its upper plateau are significantly lower than for sputtered contacts, the difference being more than one decade. This finding unequivocally reveals the non-intrinsic contact-related origin of the colossal dielectric constants in LFO. Moreover, the high-temperature step of $\varepsilon'(T)$ considerably shifts to a different frequency when the contact is varied, again demonstrating the non-intrinsic nature of the observed relaxation effects, which clearly are of MW type.

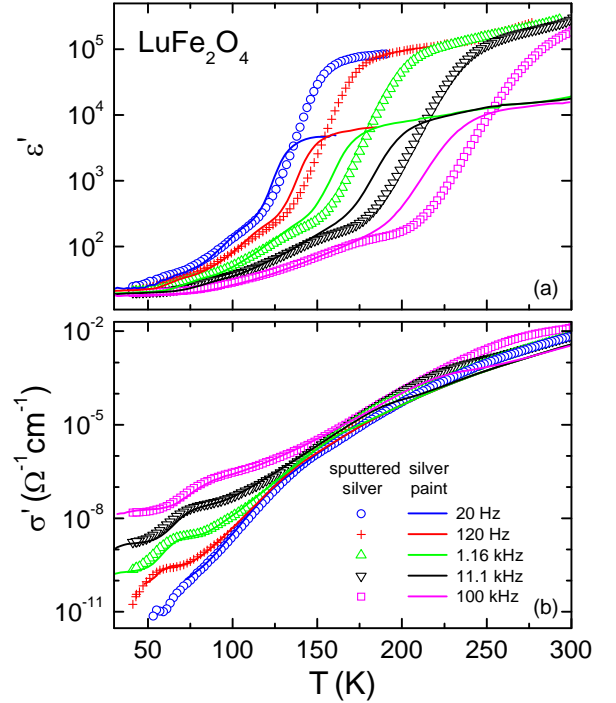

**FIG. 3.** Temperature dependence of the dielectric constant (a) and conductivity (b) of LFO measured at various frequencies. The symbols and lines represent the results from two successive measurement runs using sputtered-silver and silver-paint contacts applied to the same sample.

As discussed in detail in Refs. 5 and 6, such a variation of the dielectric behavior for different contact types can be explained by the different "wetting" of the contact surface by the applied metal layer, very similar to the findings in the prominent colossal-dielectric-constant material $CaCu_3Ti_4O_{12}$ (Refs. 6,27,28,29,30). It is plausible that the colossal dielectric constant and the MW relaxation observed in LFO (Fig. 3) arise from Schottky diodes that form at the electrode-sample interface of these semiconducting samples. Compared to sputtered contacts, the formation of these diodes should be much less effective for silver paint because there the area of direct metal-semiconductor contact can be expected to be much smaller if considering the relatively large metal particles (≥ µm) suspended in the silver paint [5,6].

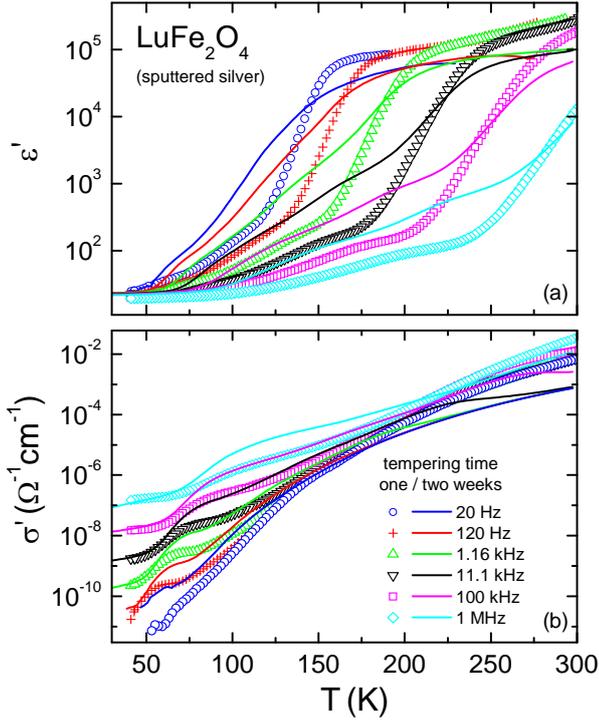

**FIG. 4.** Temperature dependence of the dielectric constant (a) and conductivity (b) of LFO measured at various frequencies. The symbols and lines represent the results obtained for two samples that were subjected to different tempering procedures and, thus, have different grain sizes as explained in the experimental section. In both cases, sputtered silver contacts were used.

Overall, the results of Fig. 3 represent a clear proof that the high-temperature relaxation and the related colossal magnitude of dielectric constant in LFO are due to electrode polarization effects. However, taking a closer look at the steplike drops of $\varepsilon'(T)$ in Fig. 3(a) reveals a significant variation of the high-temperature, but almost no variation of the low-temperature steps for the different contact types. This is further corroborated by the corresponding features in the conductivity [Fig. 3(b)]: There the typical shoulders related to the low-temperature relaxation nearly coincide for silver paint and sputtered contacts. In contrast, the corresponding features for the high-temperature relaxation clearly differ for the different contact types. Hence, the low-temperature relaxation, which leads to non-colossal but still rather high upper-plateau values of $\varepsilon'$ of the order of several hundred, nevertheless could be of intrinsic nature.

To reveal the true nature of this relaxation, the dielectric response of two samples subjected to different tempering procedures was investigated. The same type of contacts (sputtered silver) was used for both measurements. During tempering, the size of the grains within the ceramic material grows, which is revealed by the narrowing of the widths of reflexes in the X-ray diffraction patterns. Hence, any grain-boundary related contributions to the dielectric properties should differ in these two samples [5,30]. The results of those measurements are shown in Fig. 4. Indeed, both the amplitude and frequency position of the low-temperature relaxation are strongly influenced by the tempering time. Such grain-boundary related MW relaxations, leading to high dielectric constants, are well-known phenomena and also are suspected to play some role in $CaCu_3Ti_4O_{12}$ and related materials [19,30,31,32].

The amplitude of the contact-related high-temperature relaxation [upper plateau of $\varepsilon'(T)$ in Fig. 4(a)] to some extent also seems to be influenced by the grain-size variation. This is a natural effect, as the surface of the samples also should alter under tempering. For example, a slight tempering-induced variation of stoichiometry (e.g., oxygen content) at the sample surface may influence the charge carrier density and in this way affect the formation of the depletion layer of the Schottky diode. In addition, the smoothness of the sample surface also may vary with tempering [30].

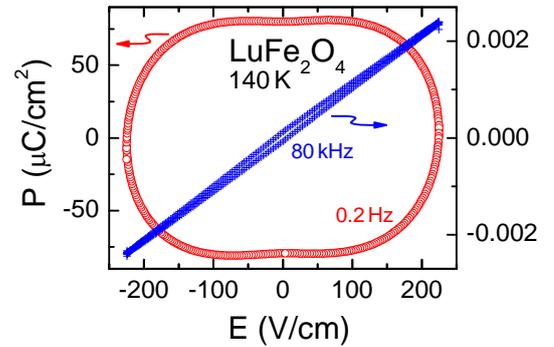

**FIG. 5.** $P(E)$ hysteresis loops of LFO at 140 K for two different frequencies.

Measuring the electric-field dependent polarization, $P(E)$, is another crucial method to check for polar order and should lead to nonlinear hysteresis curves in ferroelectrics. However, for semiconducting materials like LFO such measurements are usually hampered by charge-transport contributions. If not too large, they can be corrected by subtracting horizontal ellipses [8,33]. Figure 5 shows the conductivity-corrected $P(E)$ curves for LFO, measured at 140 K and two different frequencies. For the low frequency of 0.2 Hz, a clearly nonlinear, hysteretic response is found. However, as pointed out in [7,8], such curves can also arise from non-intrinsic effects, e.g., electrode polarization. Following the approach suggested in [8], an inspection of the dielectric results of Fig. 2 reveals that at 0.2 Hz and 140 K colossal values of $\varepsilon'$ are detected. Therefore, for this set of parameters, non-intrinsic MW effects due to electrode polarization dominate. Here a nonlinear response is expected due to the contributions of the Schottky diodes forming at the electrode-sample interfaces. In contrast, for higher frequencies, intrinsic and/or grain boundary behavior should be observed. Indeed the $P(E)$ curve at 80 kHz shown in Fig. 5 is nearly linear, clearly demonstrating the absence of any nonlinear polarization response in LFO.

Having clarified the non-intrinsic nature of the observed relaxation processes in LFO, permits an estimation of the intrinsic dc conductivity $\sigma_{dc}$ of this material from the temperature-dependent plots of the ac conductivity [Fig. 2(b)]. In this plot, $\sigma_{dc}(T)$ corresponds to the regions of weak frequency dependence as indicated by the line in Fig. 2(b) [5,30]. The latter was calculated assuming a temperature dependence $\sigma_{dc} \propto \exp[-(T_0/T)^{1/4}]$, as predicted by the variable range hopping (VRH) model [34]. The VRH model considers the phonon-assisted tunneling of localized charge carriers. $T_0$ is proportional to $\alpha^3/N(E_F)$, with $\alpha$ the inverse of the localization length and $N(E_F)$ the density of states at the Fermi level. A thermally activated behavior, $\sigma_{dc} \propto \exp[-(E/T)]$, as expected for conventional band conduction of a semiconductor, is not able to account for the experimental data. Hopping conductivity is consistent with the results of a detailed equivalent-circuit analysis of the frequency-dependent dielectric properties of single-crystalline LFO, reported in [18]. VRH was also observed in various other transition metal oxides (e.g., [30,35,36,37,38]), with values of $T_0$ ranging between $1\times10^7$ [35] and $5.4\times10^9$ K [38]. For LFO we obtain $T_0 = 3.1\times10^9$ K. The present results for $\sigma_{dc}(T)$ reasonably agree with the dc conductivity deduced from an equivalent-circuit analysis of dielectric spectra of single-crystalline LFO, which was reported in [18] for a temperature range between 200 and 400 K.

## 4 Conclusions

In summary, the presented contact and grain-size dependent dielectric investigations and polarization measurements of LFO finally prove the non-intrinsic origin of the very large dielectric constant and of the relaxation features reported for this material [9]. The main contribution to $\varepsilon'$ is provided by electrode polarization, which for sputtered contacts leads to colossal magnitudes of $\varepsilon'$ exceeding $10^5$. In addition, in the investigated ceramic samples grain-boundary effects give rise to a second MW relaxation with dielectric constants up to several 100. Overall, LFO represents a typical example of a transition-metal oxide with colossal dielectric constants generated by non-intrinsic effects, which have found considerable interest in recent years [4,5,39,40,41,42]. With an intrinsic dielectric constant of the order of 20 and a conductivity contribution due to hopping of localized charge carriers, its intrinsic dielectric properties are far from being spectacular. In particular, no dielectric signatures of polar order can be found in this material. LFO is a ferrimagnet, but it is not multiferroic.

This work was supported by the Deutsche Forschungsgemeinschaft (DFG) via the Transregional Collaborative Research Center TRR 80 (Augsburg/Munich).

———————